\documentclass[fleqn,11pt,twoside]{article}
\usepackage{espcrc1}

\usepackage{graphicx}
\usepackage[figuresright]{rotating}

\newcommand{\AmS}{{\protect\the\textfont2
  A\kern-.1667em\lower.5ex\hbox{M}\kern-.125emS}}

\title{Theoretical approaches to hadrons in nuclear matter}

\author{G. Chanfray\address{Institut de Physique Nucl\'eaire de Lyon, 
        IN2P3-CNRS et Universit\'e Claude Bernard-Lyon I, 
        F-69622 Villeurbanne cedex, France}}%
        
\begin{document}

% typeset front matter
\maketitle

\begin{abstract}
We discuss recent developments concerning the in-medium properties of hadrons 
in dense and hot matter. 
The theoretical approaches are discussed in connection with the interpretation of
experimental data from intermediate energy machines up to relativistic heavy ion
collisions. Special emphasis is put on chiral restoration  and its interplay 
with the substructure of the nucleon. 
\end{abstract}
%%%%%%%%%%%%%%%%%%%%%%%%%%%%%%%%%%%%%%%%%%%%%%%%%%%%%%%%%%%%%%%%%%%%%%%%%%%%%%%%%%%%%%%
%%%%%%%%%%%%%%%%%%%%%%%%%%%%%%%%%%%%%%%%%%%%%%%%%%%%%%%%%%%%%%%%%%%%%%%%%%%%%%%%%%%%%%%
\section{INTRODUCTION}
The problem of the behavior of hadrons in dense and hot matter is  a central
question of present day nuclear and hadronic physics. On the theoretical side
which is the main purpose of this article, a very important point is to
incorporate the relevant features of chiral dynamics into the many-body problem of
interacting hadrons to address the crucial issue of chiral symmetry restoration
with increasing baryonic density and/or temperature and its interplay with hadron
structure modification and confinement effects. The appropriate 
effective theories  have to be constrained by lattice data when it is possible 
and experimental information from a variety of sources ranging from intermediate
energy machines probing ordinary nuclear matter up to relativistic heavy ion
collisions probing hadronic matter under extreme conditions.
%%%%%%%%%%%%%%%%%%%%%%%%%%%%%%%%%%%%%%%%%%%%%%%%%%%%%%%%%%%%%%%%%%%%%%%%%%%%%%%%%%%%%%%
%%%%%%%%%%%%%%%%%%%%%%%%%%%%%%%%%%%%%%%%%%%%%%%%%%%%%%%%%%%%%%%%%%%%%%%%%%%%%%%%%%%%%%%
\section{THE ROLE OF CHIRAL SYMMETRY}
\subsection{Chiral symmetry breaking}
The $SU(2)_L\otimes SU(2)_R$ chiral symmetry of the QCD Lagrangian is certainly 
a crucial key for the understanding of many phenomena in low energy hadron physics.
This symmetry originates from the fact that the QCD Lagrangian is almost invariant 
under the separate flavor $SU(2)$ transformations  of left-handed $q_L=(u_L,d_L)$
and  right-handed $q_R=(u_R,d_R)$ light quark fields $u$ and $d$.
The explicit violation of chiral symmetry is governed by the quark mass 
$m_q=(m_u+m_d)/2\le$ 10 MeV which is much smaller than typical hadron masses 
of order $1$ GeV, indicating that the symmetry is excellent. It is however well established 
that the QCD vacuum does not possess the symmetry of the Lagrangian {\it i.e.}, chiral
symmetry is spontaneously broken (SCSB) as it is evidenced by a set of remarkable properties. 
The first one is the building-up of a
chiral quark condensate~: $\langle \bar q q \rangle =\langle \bar u u +\bar d d \rangle/ 2$
which mixes, in the broken vacuum, left-handed and right-handed
quark ($\langle \bar q q \rangle =\langle \bar q_L q_R + \bar q_R q_L\rangle/2$).
Another order parameter at the hadronic scale is the pion decay constant
$f_\pi=94$ MeV which is related to the quark condensate by the
Gell-Mann-Oakes-Renner relation~: $-2 m_q <\bar q q>_{vac}=m^2_\pi f^2_\pi$ 
valid to leading order in the current quark mass. It leads to 
a large negative value $<\bar q q>_{vac}\simeq$ -(240 MeV)$^3$ 
indicating a strong dynamical breaking of chiral symmetry. The second feature is the
appearance of soft Goldstone bosons to be identified with the almost massless pions.
Finally the chiral asymmetry of the vacuum associated with the condensation of
quark-antiquark pairs is directly visible at the level of the the hadronic
spectrum~: there is no degeneracy between possible chiral partners such as 
$\rho(770)-a_1(1260)$ or $\pi(140)-\sigma(400-1200)$. This is really true only
for the low mass part of the spectrum but higher in the spectrum  chiral symmetry
is restored since in heavy hadrons the valence quarks having higher momenta simply
decouple from the condensate. In other words, the chiral asymmetry of the vacuum
does not affect the valence quarks and the underlying plausible picture for heavy
hadrons is a rotating QCD string with bare quarks at the end-points \cite{G02}.
                %%%%%%%%%%%%%%%%%%%%%%%%%%%%%%%%%%%%%%%%%%%%%%
\subsection{Chiral symmetry restoration and hadron structure}

When hadronic matter is heated and compressed chiral symmetry is also  expected to be
restored. This picture is supported by lattice simulation showing a sharp
decrease of the quark condensate around  $T_c=170$ MeV accompanying the
deconfinement transition. However far before the critical region, partial
restoration should occur through the simple presence of hadrons. Quantitatively 
from the equation of state ({\it i.e.},  from the knowledge of the thermodynamical 
potential $\Omega(\mu_B,T)$), it follows that each hadron present with the scalar 
density $\rho_s^h (\mu_B, T)$ contributes to the dropping of the quark density through 
a characteristic quantity $\Sigma_h$ (the sigma commutator of the hadron) according to
\cite{W02}~:
\begin{eqnarray}
{\langle\langle\bar q q\rangle\rangle (\mu_B, T)\over
\langle\bar q q\rangle_{vac}}  &=&1\,+\,
{1\over 2\,\langle\bar q q\rangle_{vac}}\,{\partial\Omega(\mu_B,T) \over \partial m_q}
=1\,-\,\sum_h\,{\rho_s^h(\mu_B, T)\,
\Sigma_h\over f^2_\pi\,m^2_\pi}\nonumber\\
\Sigma_h=m_q{\partial M_h\over\partial m_q}&=& m^2_\pi\,{\partial M_h\over\partial m^2_\pi},
\qquad
\rho_s^h(\mu_B, T)={\partial\Omega(\mu_B,T) \over \partial M_h}
\end{eqnarray}
Fortunately only the lightest hadrons actually contribute since the heavy hadrons just
decouple and their corresponding sigma terms are very small. There is a very
interesting interplay between chiral restoration and hadron  structure 
since the in-medium chiral condensate behavior is driven, 
beside the influence of the equation of state,
by  the sigma term, itself related to the hadron mass obtainable in
principle  from  the lattice. Even if lattice calculations are presently 
not feasible for quark mass smaller than $60$ MeV or equivalently pion mass smaller 
than $400$ MeV,  a well controlled chiral model (possessing the correct light and
heavy quark limits) for the pion loop correction can be utilized. In the case of
the nucleon, one obtains an improved  fit with lattice data which accurately extrapolates 
to the correct nucleon mass in the physical light quark sector \cite{LTTW00}.  
This sort of interplay between  phenomenological models and QCD itself has just begun 
and will be certainly crucial in the future of this field.
                     %%%%%%%%%%%%%%%%%%%%%%%%%%%%%%%%%%%%%%%%%%%%%%%%%%%%%%
\subsection{Chiral symmetry restoration in hadronic matter}

The restructuring of the QCD vacuum associated with chiral restoration should
be visible, as  is usual in the many-body problem, at the level of the excitation
spectrum. Hence  the observable consequences should be studied looking at the 
in-medium hadronic spectral functions defined as 
$\rho_h(\omega\, ,\, \vec{q})= -(1/\pi)\, Im D_h(\omega\, ,\,\vec{q})$
where $D_h(\omega\, ,\,\vec{q})$ is a propagator or a correlator between two
currents or fields having the quantum numbers of the hadron $h$ under consideration.
Numerous works have been devoted to the centroids of the mass distributions {\it
i.e.}, the in-medium masses, but the link between the evolution of the
condensate and the masses cannot be an absolute one.  More generally the 
modification of the hadronic spectral functions is certainly not restricted to the shift of 
the centroids of  the mass distributions and we have to study the evolution of the whole shape 
of the spectral functions. Most of the time we expect a broadening of the spectral 
function (antikaons \cite{RO00,LKK02},  rho meson \cite{RW00,CS93}) 
and a dissolving of the resonances originating from 
various many-body mixing effects with the surrounding hadrons. There is however the  
notable exception of the  sigma meson for which we expect a softening and a sharpening
\cite{SC88,HA99}. The key question is the relationship between the observed reshaping 
and chiral symmetry  restoration. One possible strategy to obtain this crucial connection 
 is to make a  simultaneous study of the spectral functions associated with 
 chiral partners. A very important example is the rho meson and the axial-vector meson  
$a_1$ and it has been established
that there is a mixing of the associated current correlators trough the presence of 
the pion scalar density \cite{DEI90,CDER99}. 
                          %%%%%%%%%%%%%%%%%%%%%%%%%%%%%%%%%
\subsection{Chiral effective field theory}
The light scalar-isoscalar meson which is usually called the sigma meson is a 
particle representing the
amplitude fluctuation of the chiral order parameter $\langle\bar q q\rangle$ around the
minimum of the effective potential. The pion corresponds to the other fluctuation of the
condensate namely the phase fluctuation. It is (almost) massless since it is a mode
moving at the bottom of the ``Mexican hat'' effective potential {\it i.e.},  the so-called 
chiral circle. 
Consequently the sigma meson gets a large decay width into two-pion which makes this
object so elusive in the vacuum. In other words, the physical sigma state in QCD can be 
at most a broad resonance which is a superposition of the ur-sigma (genuine amplitude
fluctuation of the order parameter) and a two-pion state. 
To have some insight on the manifestions of chiral symmetry restoration and to make
predictions, one has to rely on effective theories which should incorporate 
the symmetry properties of the underlying QCD Lagrangian. The sigma and pion 
 taken as  effective degrees of freedom are introduced through a $2\times 2$ matrix 
$W=\sigma + i\vec\tau\cdot\vec\pi$. An alternative formulation of the resulting linear
sigma model is obtained by going from cartesian to polar coordinates according to~:
\begin{equation}
W=\sigma\, + \,i\vec\tau\cdot\vec\pi=S\,U=(f_\pi\,+\,s)\,exp
\left({i\vec\tau\cdot\vec\phi\over f_\pi}\right)
\end{equation}
The new pion field $\vec\phi$ corresponds to an orthoradial soft mode which is
automatically massless (in the absence of explicit CSB) since it is associated with
rotations on the chiral circle without cost of energy. The new sigma meson field $s$ 
describes a radial mode associated with the fluctuations of the ``chiral radius''.
As demonstrated in \cite{CEG01}, this chiral invariant $s$ field can be 
identified with the famous ``sigma meson'' 
of the relativistic QHD theories, thus giving a solution to the long-termed problem 
of their chiral status. To leading order in density and temperature, the dropping of
the chiral condensate is given by~:
\begin{equation}
R={\langle\langle\bar q q\rangle\rangle(\rho, T)\over  
\langle\bar q q\rangle_{vac}}\,
=\,{\langle\langle\sigma\rangle\rangle(\rho, T)\over  
f_\pi}\simeq1\,-\,
{\langle \phi^2\rangle (\rho, T)\over 2\, f^2_\pi}
\,-\,{| \langle s \rangle |(\rho, T)\over f_\pi}.
\end{equation}
These two contributions to the dropping of the chiral condensate yield  very different 
observable manifestations of chiral symmetry restoration. The pionic fluctuation 
piece ($\sim 20\%$ dropping at $\rho=\rho_0$) is associated with axial-vector
mixing and does not contribute to the dropping of the masses. It is the scalar piece  
($\sim 20\%$ dropping at $\rho=\rho_0$) which governs the evolution of the masses
as a consequence of the shrinking of the chiral radius.
%%%%%%%%%%%%%%%%%%%%%%%%%%%%%%%%%%%%%%%%%%%%%%%%%%%%%%%%%%%%%%%%%%%%%%%%%%%%%%%%%%%%%%
%%%%%%%%%%%%%%%%%%%%%%%%%%%%%%%%%%%%%%%%%%%%%%%%%%%%%%%%%%%%%%%%%%%%%%%%%%%%%%%%%%%%%%
\section{PIONS IN NUCLEAR MATTER}
With the above degrees of freedom, one can build an effective Lagrangian and first
address the question of the pion properties in nuclear matter. Such a Lagrangian will
depend on parameters constrained by phenomenology and data such as the very recent
accurate measurement showing that the pion-nucleon isoscalar scattering length is
compatible with zero \cite{S01}. From the Gell-Mann-Oakes Renner relation, it follows that to
leading order in density, the pion decay constant squared behaves like the quark 
condensate. Recently the attention has focused on the charge exchange scattering length
$b_1$~\cite{G00,Y02,KY01,W01,F02,KKW02}. The fit to these data~\cite{G00,Y02,KY01,F02} 
suggests a large enhancement associated with chiral restoration of the isovector scattering 
length $b_1$ in nuclei, as anticipated in ~\cite{W01}. Within the approach described 
above, one obtains~: 
\begin{equation}
{b_1^*\over b_1}=
1\,+\,{\Sigma_N\,\rho\over f^2_\pi\,m^2_\pi}
\,-\,{7\over 6}\,{\Sigma_N^{(\pi)}\,\rho\over f^2_\pi\,m^2_\pi}\simeq
\,1\,+\,0.18\,{\rho\over\rho_0}\label{B1}.
\end{equation}
which comes from the combined effect of the energy dependence \cite{KKW02} 
of the pion self-energy
and of the in-medium renormalization of the charge exchange amplitude related to the
Weinberg-Tomozawa piece of the chiral Lagrangian \cite{CEO02}. The effect is more moderate 
than in the original proposal \cite{W01} 
which ignores the pion-loop correction (last term of eq. \ref{B1}).
%%%%%%%%%%%%%%%%%%%%%%%%%%%%%%%%%%%%%%%%%%%%%%%%%%%%%%%%%%%%%%%%%%%%%%%%%%%%%%%%%%%%%%
%%%%%%%%%%%%%%%%%%%%%%%%%%%%%%%%%%%%%%%%%%%%%%%%%%%%%%%%%%%%%%%%%%%%%%%%%%%%%%%%%%%%%%
\section{SCALAR-ISOSCALAR MODES AND FLUCTUATIONS OF THE CHIRAL CONDENSATE}

As mentioned before, the genuine  sigma meson is defined as the quantum fluctuating mode
of the chiral order parameter around the minimum of the effective potential. In nuclear matter 
the minimum is shifted   and it is  crucial to address this question
of matter stability in chiral effective theories. The energy density taken as a function 
of the order parameter, which we call generically $\langle\ S\rangle$, is  the appropriate 
effective potential~:
\begin{equation}
\epsilon(\rho,\langle S\rangle)=\sum_{p<p_F}\,\sqrt{ p^2\,+\,M^*_N(\langle S\rangle)}\,
+\,V(\langle S\rangle)\,+\,C_V\,\rho^2\,.
\end{equation}
$V(\langle S\rangle)$ is the ``Mexican hat'' potential generating vacuum symmetry
breaking and the last term corresponds to some vector repulsion. In the NJL model the
nucleon can be built as a quark-diquark bound state and the constituent quark mass 
$M_q^*=g_q\,\langle S\rangle$ plays the
role of the order parameter \cite{BT01}. 
In the chiral version of relativistic theories of the
Walecka type \cite{CEG01}, the sigma field can be identified  with the chiral 
invariant scalar field
from which one can dynamically generate the effective nucleon mass 
$M^*_N=g_{ SN}\,\langle\ S\rangle$. 
Independently  of the particular chiral model, the decrease of the curvature of the Mexican 
hat potential, as one moves away from the vacuum state, implies an increase 
of the attraction between nucleons. 
The needed extra-repulsion to get matter stability can be obtained 
if the sigma-nucleon coupling constant 
$g^*_{SN}$ becomes a decreasing function of the order parameter. This is precisely 
what happens if an infrared cut-off simulating confinement is incorporated in the NJL model
\cite{BT01}  and the same effect appears in the QMC model from the polarization of the 
confined quark wave  functions \cite{G88}. 
Since the coupling constant  $g^*_{SN}=\partial M^*_N/\partial \langle\ S\rangle$
is essentially the scalar response of the nucleon, which is in principle calculable on
the lattice, one gets a connection between nuclear saturation and QCD itself! The sigma
mass is defined as the curvature of the effective potential $m^2_\sigma=
\partial^2\epsilon/\partial\sigma^2$. In case of structureless nucleons, it strongly
decreases with the condensate as a consequence of chiral restoration.  However when 
nucleon polarization and/or confinement effects are  included, the sigma mass 
remains remarkably stable \cite{BT01,CE02BIS}. 
This feature has a direct consequence on the scalar 
susceptibility defined in terms of the correlator of the quark density fluctuations 
which encodes the properties of the sigma meson~:  
\begin{equation}
\chi_s={\partial\langle\langle \bar q\, q\rangle\rangle\over \partial
m}=-\,i\,
\int \,dt\,d{\vec r}\,\langle\langle \delta\,\bar q\, q (0 , 0) , \,
\delta\,\bar q\, q ({\vec r} , t)\rangle\rangle.
\end{equation}
Such a quantity can be calculated on the lattice at finite temperature. It becomes very
large near the phase transition as it should since the fluctuations of the order
parameter become large as   a second- or a weak first-order phase
transition is approached. In addition lattice calculations also show that
the pseudoscalar susceptibility (pionic channel) becomes identical with the scalar one
beyond the transition point indicating chiral restoration \cite{KA02}. At finite density  
since lattice data are  scarce, one has to rely on models. 
According to a calculation within the kind of effective theory described above, the 
convergence of these two susceptibilities at density larger than $3\rho_0$ seems to
occur. In this approach  the scalar fluctuations are transmitted by the sigma meson and 
are relayed by the nucleons \cite{CE02}.

The in-medium scalar-isoscalar modes in nuclei have been studied experimentally through
two-pion production in reactions induced either by pions \cite{B96,ST00} or photons 
\cite{M02} on various nuclei.
All these collaborations have observed a systematic A dependent downwards shift of the
strength when the pion pair is produced in the sigma meson channel as it is apparent
on the photoproduction data of the TAPS collaboration \cite{M02} (see right panel of Fig. 1).
So far, conventional
calculations are not able to reproduce the data and the  proposed explanations rely 
on a in-medium modification of the unitarized $\pi\pi$ 
interaction in the final state of the reaction. Historically the first advocated 
medium effect is related to the softening of the pion dispersion hence modifying the two-pion 
propagator in the unitarized $\pi\pi$ interaction \cite{SC88,VO99}. The other effect 
proposed by Hatsuda and Kunihiro is related to 
chiral restoration through the dropping of the sigma mass and the scalar 
condensate \cite{HA99,ADZC00}. This is a very general argument since chiral restoration 
implies a softening of a collective scalar-isoscalar mode, the sigma meson, which becomes
degenerate with its chiral partner, the pion, at full restoration. This implies that
at some density the sigma mass is twice the pion mass and  a sharp
cusp-structure near threshold is generated. This spectral enhancement \cite{HA99}  is actually
intimately related to enhanced in-medium fluctuations of the chiral condensate. 
This effect is however weakened if the nucleon polarization effect is included \cite{CE02BIS}. 
Consequently at moderate density (see Fig. 1) corresponding to the actual experimental 
situation, the medium effect is mainly driven by pions. Indeed a detailed model
calculation including only this latter medium effect \cite{ROV02} 
is able to reproduce quite decently
the photoproduction data of the TAPS collaboration, as shown on Fig. 1. 
Hence whatever the vacuum nature
of the sigma meson and the medium effect affecting it are, it appears that 
the sigma pole moves at lower energy with reduced width from in-medium chiral dynamics.
\begin{figure}[h]\vspace*{-1cm}
\begin{minipage}[b]{0.40\textwidth}
%\begin{figure}[h]
\begin{center}
\includegraphics[width=5 cm,height=5cm]{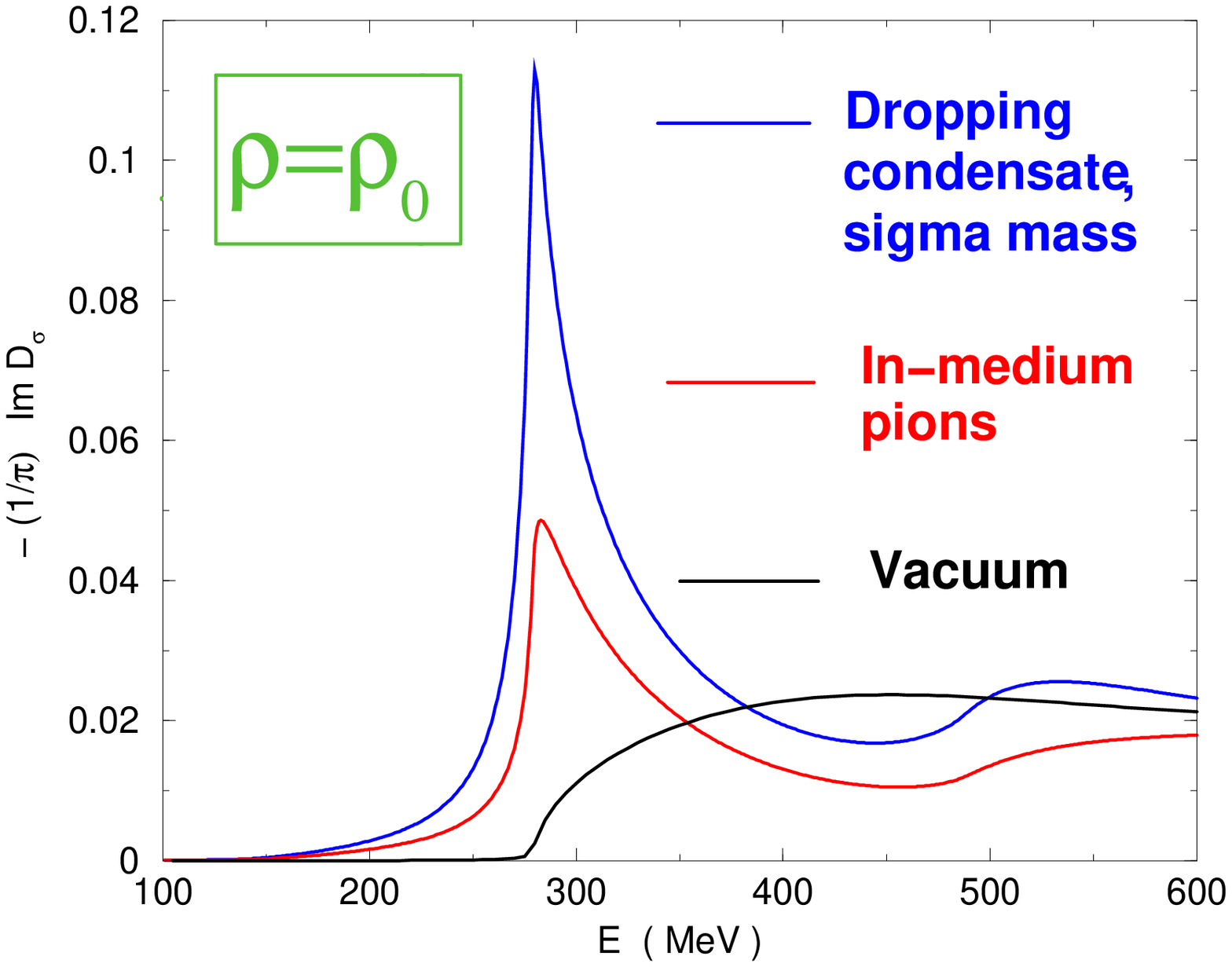}
\end{center}
%\end{figure}
%\begin{figure}[h]
\begin{center}
\includegraphics[width=5 cm,height=5cm]{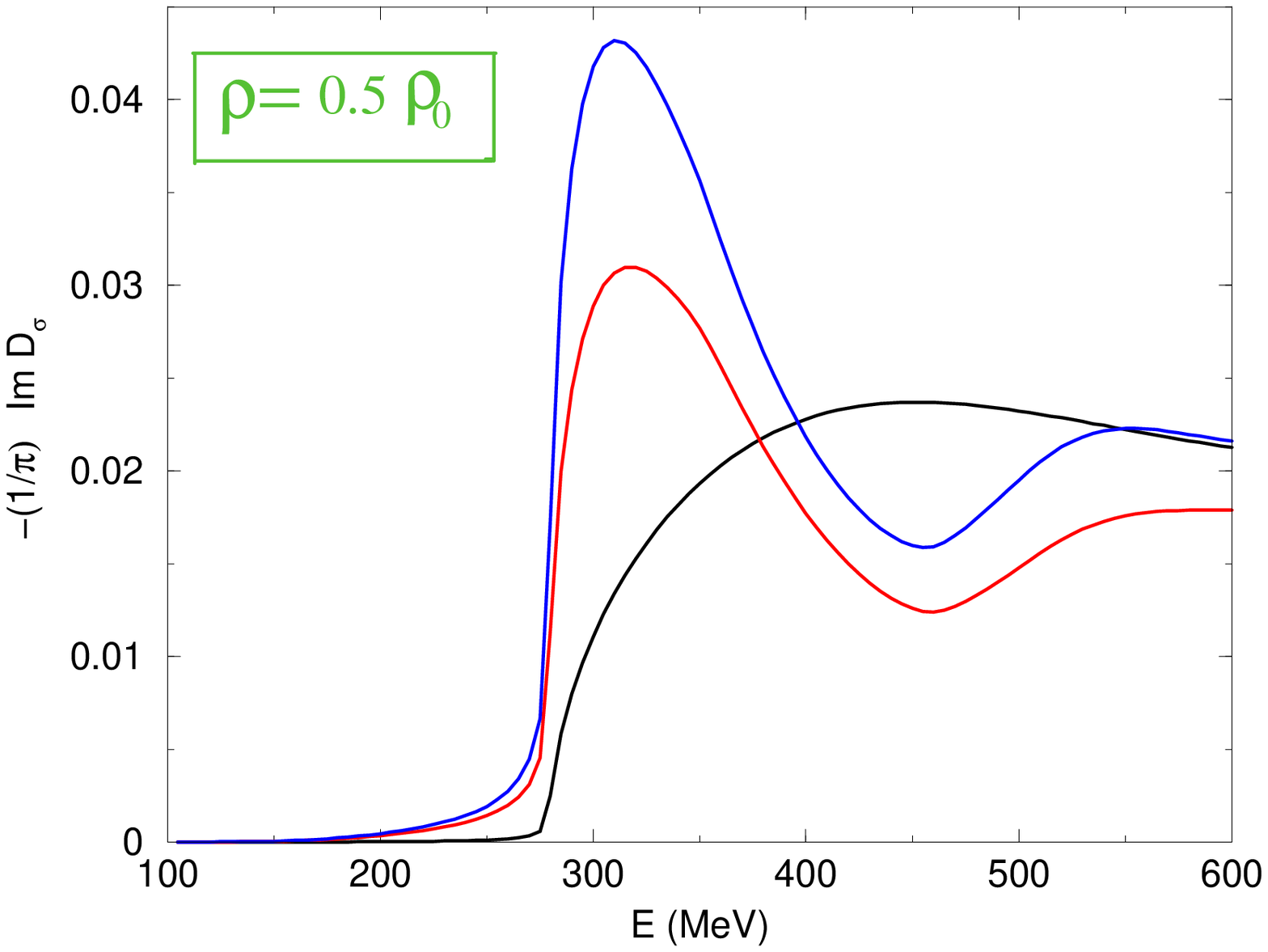}
\end{center}
\vspace{0.3 cm}
\end{minipage}
\hfill
\begin{minipage}[b]{0.60\textwidth}
\begin{center}
\includegraphics[width=9 cm,height=12 cm]{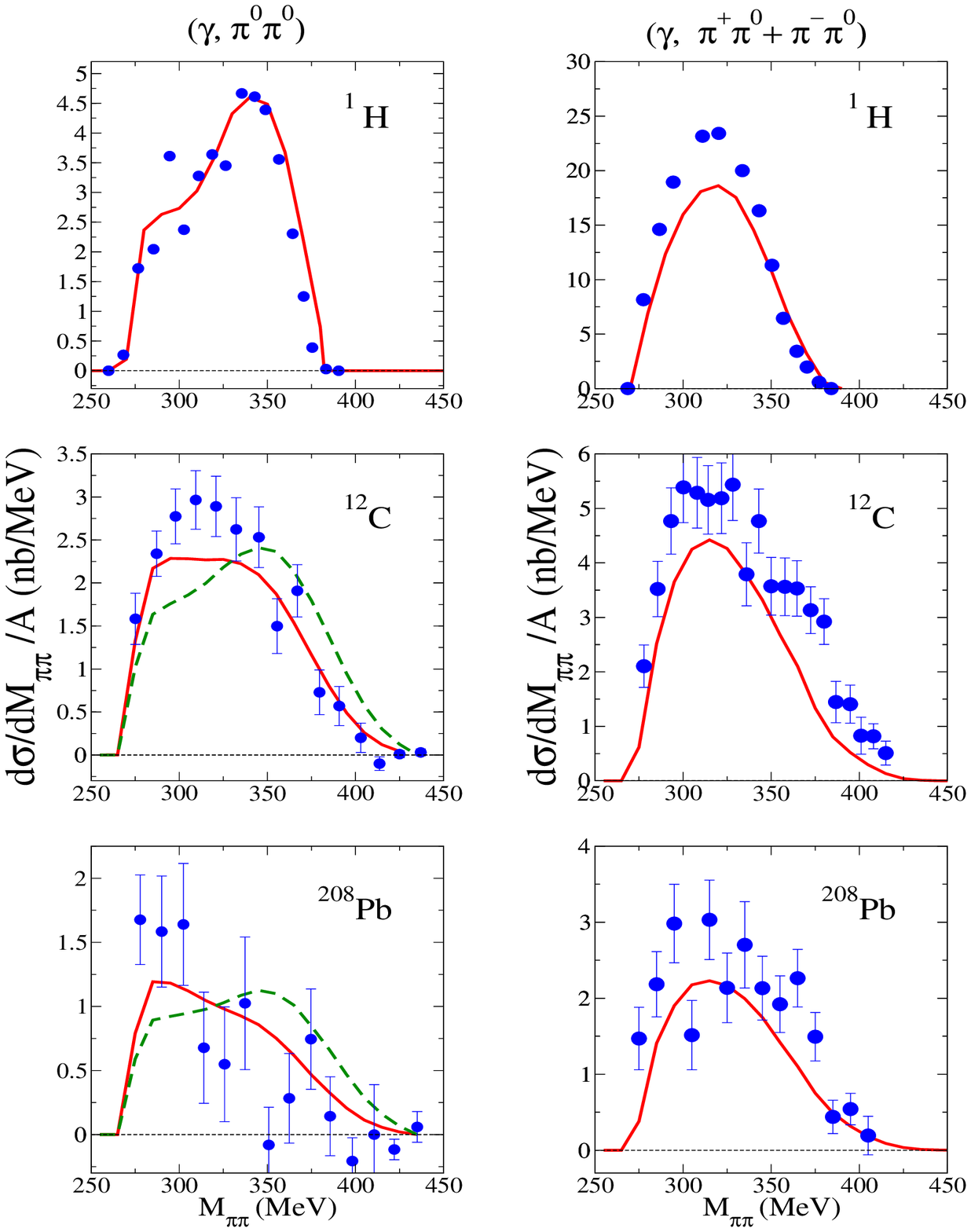}
\end{center}
\end{minipage}
\vspace*{-1cm}
\caption{Left panel: sigma strength function in the vacuum (lower curves), 
with in-medium pionic effects (middle curves) and with chiral restoration on top of 
in-medium pionic effects 
(higher curves). Right panel: two-pion invariant mass distributions  for $\pi^0 \pi^0$
and $\pi^\pm \pi^0$ photoproduction on various nuclei \cite{M02}
compared with a calculation with in-medium pionic  effects (full lines) and without 
(dashed  lines) in the final  $\pi\pi$ interaction \cite{ROV02}.}
\end{figure}
\begin{figure}[h]
\vspace*{-2 cm}
\begin{minipage}[c]{0.5\textwidth}
\vspace*{0.5cm}
\includegraphics[width=0.9\textwidth]{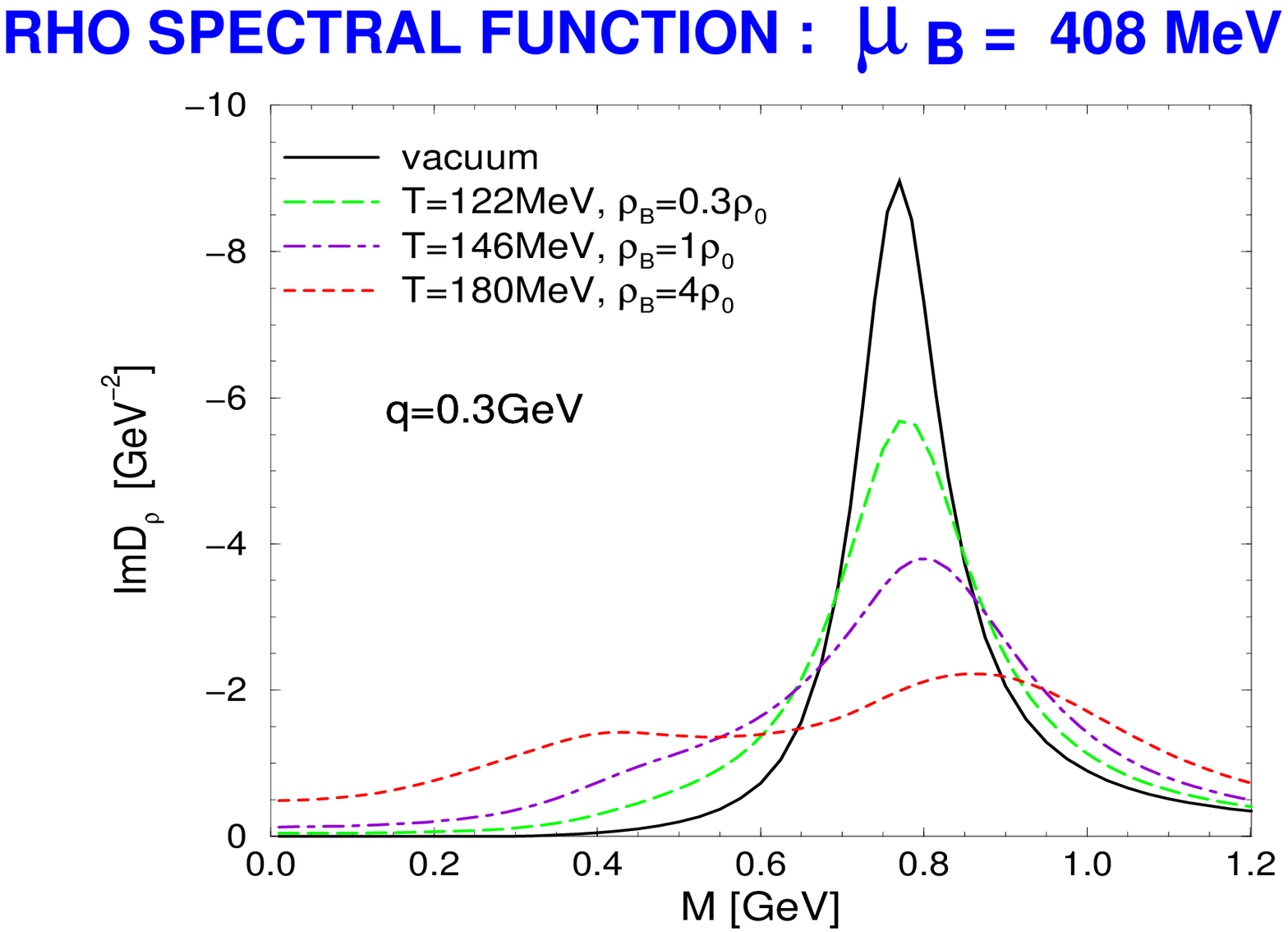}
\end{minipage}
\hfill
\begin{minipage}[c]{0.5\textwidth}
\includegraphics[width=0.8\textwidth]{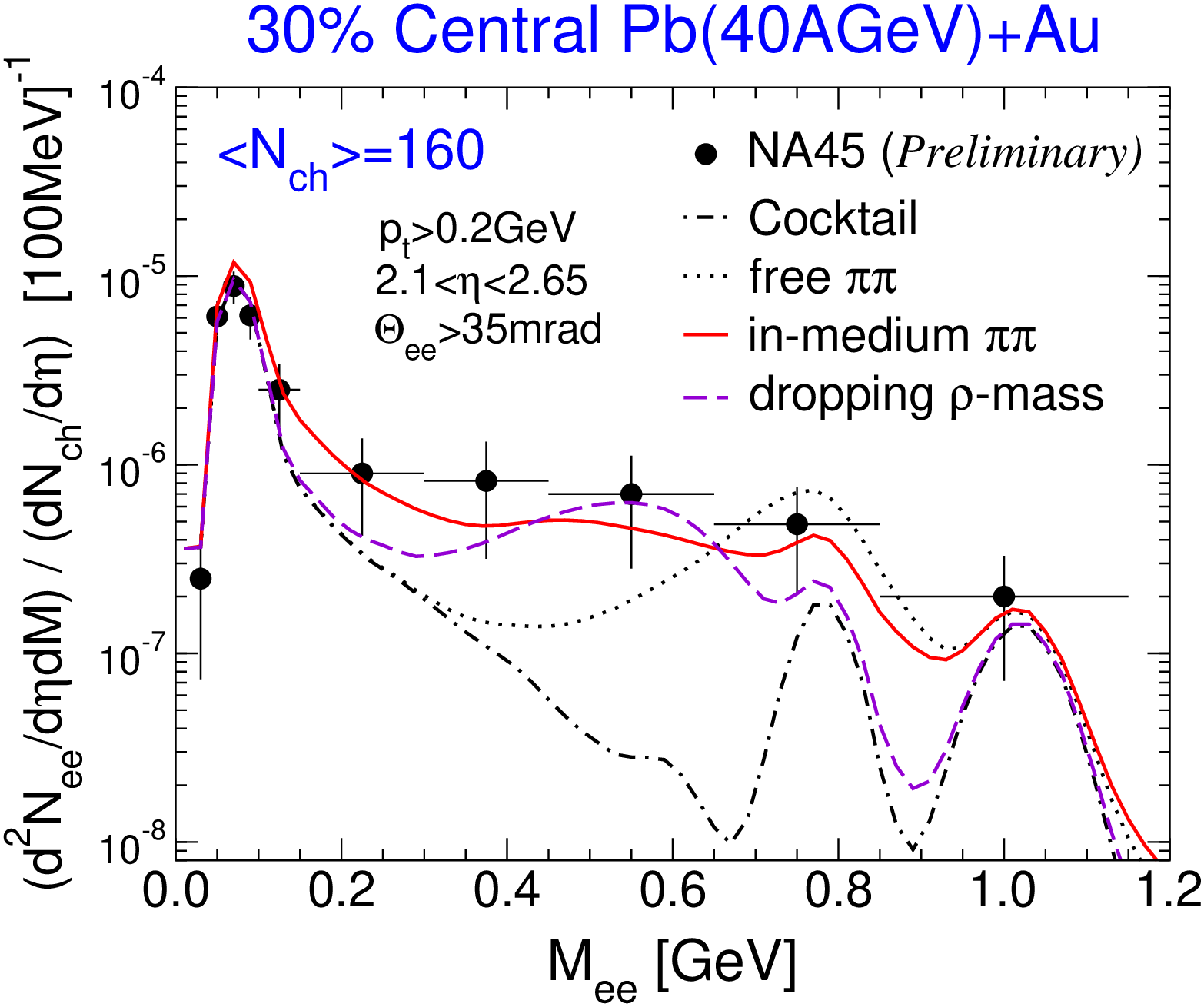}
\end{minipage}
\vspace*{-1cm}
\caption{Left panel~: rho meson spectral function at fixed baryon chemical potential 
$\mu_B=408$
GeV and various temperatures \cite{RW00};
Right panel~:  thermal fireball calculations \cite{RAPP02} plus hadronic decay cocktail
compared to the CERES data at low SPS energy \cite{A01}.}
\end{figure}

%%%%%%%%%%%%%%%%%%%%%%%%%%%%%%%%%%%%%%%%%%%%%%%%%%%%%%%%%%%%%%%%%%%%%%%%%%%%%%%%%%%%%%
%%%%%%%%%%%%%%%%%%%%%%%%%%%%%%%%%%%%%%%%%%%%%%%%%%%%%%%%%%%%%%%%%%%%%%%%%%%%%%%%%%%%%%
\section{TOWARDS HIGH BARYONIC DENSITIES AND PERSPECTIVES}
What happens at higher density is largely unknown and it is a major experimental 
challenge to investigate this high density region which should complement the high 
temperature domain studied at RHIC or at CERN/SPS. In that respect, there is a
GSI project to study compressed baryonic matter matter with heavy ions in the range 
$10-40$ AGeV. One crucial point is of course the nature of the phase transition.
For instance lattice data show that at finite temperature deconfinement and chiral
restoration occur simultaneously but
we do not really understand the reason and one question is whether this feature survives in
the high density transition. Even if this transition is very difficult to reach
experimentally one can adopt a pragmatical bottom-to-top attitude looking at, beside
the usual macrodynamical variables (flow) probing the eos, the behavior of the
hadron spectral functions as precursor effects of the phase transition. This opens 
many theoretical challenges which can be illustrated with dilepton production. It is
well known that a strong excess  of  dileptons  below the rho
mass has been observed at CERN/SPS and it has been attributed to strong medium effects 
in the interacting fireball. 
The dilepton production is actually related to the imaginary part of the
current-current correlation function in the vector channel.
According to the well established vector
dominance phenomenology, this correlator is accurately saturated by the rho meson
at least for the radiation from the interacting fireball before freeze out. Hence
dilepton production probes the rho meson spectral function in the dense and hot medium.  
Experimentally one sees a 
broadening and a flattening of the rho meson spectral function. The resulting spectrum
is very similar in shape with what would be radiated by a perturbative quark-gluon
phase in which chiral symmetry is restored \cite{RW00}. 
The DPR has been studied using various
methods ranging  from density expansions up to transport codes together with 
model independent in spirit theoretical tools for calculating or constraining 
the current-current correlator such as chiral reduction formalism, QCD sum rules or
Weinberg sum rules. One important  very much debated question concerns the mechanism
associated with chiral restoration. In the Rho-Brown dropping scenario \cite{BR91}, the rho
meson mass plays the role of an order parameter. Another mechanism is the
axial-vector mixing. The emission and the absorption of thermal (finite temperature) 
or virtual (finite density)  pions in the medium is able to transform a 
vector current  into an  axial current and at
full restoration the axial and vector correlators become identical \cite{DEI90,CDER99}. 
This mixing effect together with other ones such as the mixing of the rho
with some $N^*-h$ configurations has been explicitly incorporated in the very
detailed hadronic calculation \cite{RW00} of the in-medium rho spectral function (see fig. 2). 
Once the local rate is space-time
evolved through a realistic fireball expansion, this approach (full curve on the left panel of
fig.2)  correctly reproduces the CERES data and in particular the most 
recent data at lower SPS  energy (40 AGeV) for which the low mass enhancement effect is even
more important. This is compatible with the fact   that the medium effects  are  
mainly driven by baryons. 
Such a study has to be pursued in direction of higher density  with the
forthcoming HADES data in the GeV range and the future project at GSI. However
on the theoretical size
some questions remain to be clarified such as the chiral status of the $N^*(1520)$
and this again might be related to an interplay between chiral dynamics
and nucleon structure. In addition  some new theoretical suggestions have been
proposed such as the simultaneous softening of the rho  and sigma mesons \cite{YHHK02} 
or the fate of vector dominance at finite temperature or density. 
In a set of recent works mainly driven by Harada and Yamawaki \cite{HY01} 
a novel way of matching an 
effective field theory with the underlying QCD in the sense of a Wilsonian 
renormalization group equation has been proposed. This has been explicitly worked 
out in the HLS model  where the rho meson is generated as the gauge boson of a hidden
local symmetry. By matching the axial and vectors correlators at a well chosen scale
$\Lambda\sim 1$ GeV, the bare parameters of the effective Lagrangian are determined. 
Using the renormalization group equations   with inclusion of quadratic divergences, 
the physical quantities such as masses
and coupling constants are then obtained. It has been found that the usual vector dominance 
is recovered in the vacuum but is lost with increasing temperature and density. One
obtains a vector manifestation of chiral symmetry in which the (longitudinal) rho
becomes massless at the chiral phase transition point.
This kind of approach which allows to constrain effective theories is certainly very
encouraging and has to be pursued with the explicit incorporation of other degrees
of freedom, and in first rank the scalar-isoscalar degrees of freedom. It is also clear
that the deep understanding of hot and dense hadronic matter is a vast theoretical
challenge. It certainly necessitates the interplay between model building, lattice QCD
at finite chemical potential and non-perturbative methods such as renormalization
group equations or methods issued from the many-body problem. 

\smallskip\noindent 
Acknowledgements. I am grateful to M. Ericson for critical comments on the manuscript.

\end{document}